\documentstyle[psfig]{mn}
\newif\ifAMStwofonts

\ifoldfss
  \ifCUPmtlplainloaded \else
    \NewTextAlphabet{textbfit} {cmbxti10} {}
    \NewTextAlphabet{textbfss} {cmssbx10} {}
    \NewMathAlphabet{mathbfit} {cmbxti10} {} 
    \NewMathAlphabet{mathbfss} {cmssbx10} {} 
  \fi
  \ifAMStwofonts
    \ifCUPmtlplainloaded \else
      \NewSymbolFont{upmath} {eurm10}
      \NewSymbolFont{AMSa} {msam10}
      \NewMathSymbol{\upi}     {0}{upmath}{19}
      \NewMathSymbol{\umu}     {0}{upmath}{16}
      \NewMathSymbol{\upartial}{0}{upmath}{40}
      \NewMathSymbol{\leqslant}{3}{AMSa}{36}
      \NewMathSymbol{\geqslant}{3}{AMSa}{3E}

      \let\leq=\leqslant 
       
    \fi
  \fi
\fi 

\ifnfssone
  \newmathalphabet{\mathit}
  \addtoversion{normal}{\mathit}{cmr}{m}{it}
  \addtoversion{bold}{\mathit}{cmr}{bx}{it}
  \newmathalphabet{\mathbfit} 
  \addtoversion{normal}{\mathbfit}{cmr}{bx}{it}
  \addtoversion{bold}{\mathbfit}{cmr}{bx}{it}
  \newmathalphabet{\mathbfss} 
  \addtoversion{normal}{\mathbfss}{cmss}{bx}{n}
  \addtoversion{bold}{\mathbfss}{cmss}{bx}{n}
  \newmathalphabet{\mathbfit} 
  \addtoversion{normal}{\mathbfit}{cmr}{bx}{it}
  \addtoversion{bold}{\mathbfit}{cmr}{bx}{it}
  \newmathalphabet{\mathbfss} 
  \addtoversion{normal}{\mathbfss}{cmss}{bx}{n}
  \addtoversion{bold}{\mathbfss}{cmss}{bx}{n}     
  \ifAMStwofonts
    \ifCUPmtlplainloaded \else
      \UseAMStwoboldmath
      \makeatletter
      \new@mathgroup\upmath@group
      \define@mathgroup\mv@normal\upmath@group{eur}{m}{n}
      \define@mathgroup\mv@bold\upmath@group{eur}{b}{n}
      \edef\UPM{\hexnumber\upmath@group}
      \new@mathgroup\amsa@group
      \define@mathgroup\mv@normal\amsa@group{msa}{m}{n}
      \define@mathgroup\mv@bold\amsa@group{msa}{m}{n}
      \edef\AMSa{\hexnumber\amsa@group}
      \makeatother
      \mathchardef\upi="0\UPM19
      \mathchardef\umu="0\UPM16
      \mathchardef\upartial="0\UPM40
      \mathchardef\leqslant="3\AMSa36
      \mathchardef\geqslant="3\AMSa3E

      \let\leq=\leqslant 

    \fi
  \fi
\fi 

\ifnfsstwo
  \DeclareMathAlphabet{\mathbfit}{OT1}{cmr}{bx}{it}
  \SetMathAlphabet\mathbfit{bold}{OT1}{cmr}{bx}{it}
  \DeclareMathAlphabet{\mathbfss}{OT1}{cmss}{bx}{n}
  \SetMathAlphabet\mathbfss{bold}{OT1}{cmss}{bx}{n}
  \ifAMStwofonts
    \ifCUPmtlplainloaded \else
      \DeclareSymbolFont{UPM}{U}{eur}{m}{n}
      \SetSymbolFont{UPM}{bold}{U}{eur}{b}{n}
      \DeclareSymbolFont{AMSa}{U}{msa}{m}{n}
      \DeclareMathSymbol{\upi}{0}{UPM}{"19}
      \DeclareMathSymbol{\umu}{0}{UPM}{"16}
      \DeclareMathSymbol{\upartial}{0}{UPM}{"40}
      \DeclareMathSymbol{\leqslant}{3}{AMSa}{"36}
      \DeclareMathSymbol{\geqslant}{3}{AMSa}{"3E}

      \let\leq=\leqslant 

    \fi
  \fi
\fi 

\ifCUPmtlplainloaded \else
  \ifAMStwofonts \else 
    \def\upi{\pi}
    \def\umu{\mu}
    \def\upartial{\partial}
  \fi
\fi

\title{The X-ray emission from  Nova V382 Velorum:\\
II. The super-soft component observed with BeppoSAX }
\author[M. Orio et al.]
        {M. Orio$^{1,2}$, A. N. Parmar$^3$, J. Greiner,$^4$
H. \"Ogelman$^5$, S. Starrfield$^6$, E. Trussoni$^1$\\
$^1$ Istituto Nazionale di Astrofisica (INAF), Osservatorio Astronomico di Torino, Strada
Osservatorio, 20, I-10025 Pino Torinese (TO), Italy\\
 $^2$ Department of Astronomy, 475
N. Charter Str., University of Wisconsin, Madison WI 53706, USA\\
 $^3$ Astrophysics Division, Space Science Dept. of ESA,
   ESTEC, Postbus 299, 2200 AG Noordwijk, The Netherlands\\
$^4$ Astrophysical Institut, 14882 Potsdam, An der
Sternwarte 16, FRG\\
$^5$ Department of Physics, University of Wisconsin, 1500 University Ave., Madison WI 53706, USA\\
$^5$ Dept of Physics and Astronomy, P.O. Box 87150,
Arizona State University, Tempe AZ  85287-1504, USA}
\date{
      Received; Accepted }

\pagerange{\pageref{firstpage}--\pageref{lastpage}}
\pubyear{1994}
\begin{document}

\maketitle

\label{firstpage}

\begin{abstract}
Nova Velorum 1999 (V382 Vel) was observed by BeppoSAX 
6 months after optical maximum and  was detected as a bright
X-ray supersoft source, with a count rate 3.454$\pm$0.002 cts s$^{-1}$
in the LECS.  It was the softest and most
luminous supersoft source observed with this instrument. The flux in the
 0.1--0.7 keV 
range was not constant during the observation.  It dropped
 by a factor of 2 in less than 1.5 hour and then
was faint for at least 15 minutes,
without significant spectral changes. 
 The observed spectrum is not well fit with
  atmospheric models of a hot, hydrogen burning white dwarf.
 This is due mainly to a
supersoft excess in the range 0.1-0.2 keV, but the fit can be significantly 
improved
at higher energy if at least one  emission feature is superimposed.
We suggest that a ``pseudocontinuum'' was detected, consisting of 
 emission lines in the supersoft X-ray range
superimposed on the thermal continuum of a white dwarf atmosphere.
As a result, an accurate determination of the effective
temperature and gravity of the white dwarf at this post-outburst
stage is not possible.

\end{abstract}


\section{Introduction}

Nova Velorum 1999 (V382 Vel) was the second brightest nova of the
 last 50 years (V=2.6) (Seargent \& Pearce, 1999).
It  was a ``fast ''
O--Ne--Mg nova, with v$\simeq$4000 km s$^{-1}$,
 t$_2$= 6 d, t$_3$=10 d (Della Valle et al. 1999, Shore
et al. 1999).        
The nova was immediately declared
a  Target of Opportunity by the BeppoSAX Mission Scientist.

 BeppoSAX carries instruments that cover the energy range
0.1-300 keV.
 The instruments used are the coalined Low-Energy Concentrator
Spectrometer (LECS;
0.1--10~keV; Parmar et al. 1997), the Medium-Energy Concentrator
Spectrometer (MECS; 1.8--10~keV; Boella et al. 1997),          
and the Phoswich Detection System (PDS; 15--300~keV; Frontera et al.
1991).
The LECS and MECS, used in this work, consist of grazing incidence
telescopes with imaging gas scintillation proportional counters in
their focal planes. 

Novae in outburst have been observed to emit
X-rays  due to thermal emission of shocked ejecta (see Orio
et al. 2001a, and references therein).
The outburst is normally due to a radiation
 pressure driven wind and not to a shock wave, however
shocks can be produced in interacting winds, or interaction
between the ejecta and the circumstellar medium. After a few months,
luminous {\it ``supersoft''} X-ray emission (luminosity
of the order 10$^{37-38}$ erg s$^{-1}$) has also been
observed. The previous detections have been attributed 
to residual hydrogen burning in a shell on the white dwarf remnant
(e.g. \"Ogelman et al. 1993, Krautter et al. 1996,
Orio \& Greiner 1999).
 We expect to detect in this case an
atmospheric continuum at T$_{\rm eff}$=20-80 eV
and the absorption edges of the white dwarf (or even {\it emission
edges} if the effective temperature is extremely high).         

 V382 Vel was observed with BeppoSAX, with ASCA and RossiXTE two weeks 
after the outburst (Orio et al. 1999a, Orio et al. 2001b,
Mukai \& Ishida 1999, 2001) as a hard X-ray source (with
plasma temperature kT$\simeq$7 keV). It cooled rapidly
in the first two months after outburst to kT$\simeq$2.4
keV (Mukai \& Ishida 2001). In a recent
paper (Orio et al. 2001b; hereafter,
Paper I) we attributed the initial X-ray emission to shocks
in the nebula.  Since the initially very large 
intrinsic absorption of the ejected nebula was thinning
out, the equivalent N(H) decreased rapidly 
 (Mukai \& Ishida 2001). Thus, in the second BeppoSAX
observation we hoped to detect the super-soft X-ray emission with
the LECS and derive useful information on the nature of the
 white dwarf.   The peak temperature of the hot white dwarf remnant,
and the absorption edges that indicate the underlying chemical
composition, are extremely important in order to constrain the
physical models. In addition to this, the length of the
constant bolometric luminosity phase is an essential parameter
in order to understand whether the nova retains accreted mass after the
outburst and  is therefore a candidate type Ia supernova
(or, even, a candidate neutron star formed by accretion induced
collapse). 


\section{The supersoft X-ray source}
 
In the second BeppoSAX observation 
on 1999 November 23, the actual length of the LECS
exposure was 12.4 ksec, over about a total 16 hours in 11 time intervals, 
lasting for t$\leq$1200 sec each. The MECS
exposure  time was 25.9 ksec during the same 16 hours.   The count rate measured with the
LECS was extremely high, 3.4540$\pm$0.0021 cts s$^{-1}$ in the 0.1-4.0 keV range,
due to the emergence of  supersoft X-ray emission 
 (Orio et al.  1999b). This high count rate was not unexpected,
and it would be equivalent (assuming for simplicity
 a blackbody at 40 eV and N(H)=2 $\times$ 10$^{21}$ cm$^{-2}$)
to  $\approx$50 cts s$^{-1}$ with the ROSAT PSPC.  A count rate of
75 cts s$^{-1}$ was measured for V1974 Cyg
at maximum with the PSPC (Krautter et al. 1996).
 In the 0.8-10.0 keV range the count rate was only 0.1030$\pm$0.0038, 
consistent with the MECS count rate
0.0454$\pm$0.0015 cts s$^{-1}$ (more than a factor of 3 lower than in 
June of 1999).  
There was no PDS detection  with a 2$\sigma$ upper limit of 
0.080 cts s$^{-1}$ in the 15-50 keV range.
 We have already discussed the evolution of the hard  X-ray emission
(Paper I). We found that more than one component
was necessary to fit the LECS spectrum, however we also
 concluded that there was no component with a 
higher plasma temperature than $\simeq$1 keV, that N(H) was consistent
with the interstellar value and that the supersoft portion of the flux was 
dominant.

Remarkably, the supersoft X-ray flux  (0.1-0.7 keV) was 
variable. No significant variability was detected at higher energy.
 Overall, there was irregular flickering with  time scales
of minutes, and  as Fig. 1 shows, in the 9th observation 
(after $\approx$13 hours from the beginning of the exposures) the
background-subtracted count rate decreased dramatically, 
by approximately a factor            of 2. This low state 
lasted  during the whole LECS coverage of 15 minutes, spaced about 5000 seconds 
from two observation in which the average count rate was twice higher. 
Close to the end of the $\approx$16 hours of intermittent observations, 
the count rate 
decreased again in the few minutes (see bottom panel of Fig. 1).
 If we missed other episodes of this type, they must have been shorter
than the 4000--5000 seconds that elapsed between the observations. 
  One possible explanation for the sudden decrease in count rate 
is of course that a dense clump intervened along the line of sight. However, 
the spectrum  was definitely supersoft during the whole observation
and became slightly softer during the dip. Instead,
the additional absorption of a thick clump would absorb the softer
portion of the spectrum more and produce an apparently ``harder'' spectrum.

Is this phenomenon linked with orbital variability? Given
the observed periodicity at optical wavelengths (Bos et al. 2001)
the orbital period of V382 Vel is likely to be 3.5 hours.
 The time that elapsed between minima
in the last observation span was a little over 3
hours.  We folded the LECS  lightcurve with  the optical modulation period
and even if $\approx$70\% of the phase was covered, we
 could not detect any modulation in supersoft X-ray flux. 
We also note that the semi-amplitude
of the optical modulation is only 0.02 mag, while the X-ray count rate
has a large variation.  

The time scale of the phenomenon
does not give very significant upper limits on the size
of the obscuring region: the upper limit on the size of an obscuring  
clump in the nebula, assumed to be moving at v=4000 km s$^{-1}$
for 5000 seconds, is 2 $\times 10^{12}$ cm 
(a small fraction of the nova shell radius at this epoch). Assuming 
instead some other type of phenomenon, 
connected with the possible orbital period, an upper limit on the size
is obtained assuming the speed of light for the 
obscuring source: 1.5 $\times 10^{14}$ cm. 

This variability is a truly puzzling phenomenon which we do 
not fully understand. We note that also V1494 Aql (N Aql 1999
no.2), which was observed with
Chandra and also detected as a supersoft X-ray source, showed
time variability in supersoft X-rays: a flare 
(with increase of flux by a factor 6) that lasted for about 15 minutes
and pulsations every 42 minutes (Starrfield et al. 2001). 
Even for this nova, the supersoft X-ray variability time scale was very short.

\begin{table*}
  \caption{Spectral fit parameters obtained applying two 
 NLTE models  with log(g)=8.5 and different abundances (25\% the cosmic value,
 NLTE-1, and LMC-like depleted, NLTE-2), plus {\sc MEKAL} and
 lines, to the BeppoSAX LECS and MECS
spectra of Nova Vel observed in 1999 November.
 The absorption, N(H), is in units of $10^{22}$~atom~cm$^{-2}$.
T$_{\rm eff}$ is the WD atmospheric temperature, kT the {\sc MEKAL} 
plasma temperature (of the ejecta), F$_a$ is the absorbed flux and
F$_x$ the unabsorbed flux, $\chi^2$/dof is the reduced $\chi^2$,
dof the number of degrees of freedom.} 
\begin{tabular}{|l c c c c c c c c |}              \hline
 Model & N(H) &  T$_{\rm eff}$($\times 10^5$ K) & kT (keV)&  F$_a$ & F$_x$ & $\chi^2$/dof & dof \\
\hline
 A: NLTE-1+MEKAL               & 1.38     & 5.6 & 0.896 &  & & 4.53  & 105 \\
 B: NLTE-1+MEKAL+L(0.449 keV)  & 2.03     & 4.0 & 0.801 & & & 1.70  & 102 \\
 C: NLTE-1+MEKAL+L(0.243 keV+0.449 keV & 2.07  & 4.0 & 0.836 &
 1.6 $\times$ 10$^{-9}$ &  6.1$\times$ 10$^{-6}$  & 1.67  & 96 \\
      +6.4 keV)               &          &     &        & & & \\
 D: NLTE-2+MEKAL              & 1.56 &  5.3 & 0.069 &   &  &  12.66 & 122 \\
 E: NLTE-2+MEKAL+L(0.453 keV)     &   1.98 & 5.2 & 0.826 &  &   & 1.57  & 117 \\
 F: NLTE-2+MEKAL+L(0.491 keV+6.52 keV) & 1.98 & 5.2 & 0.813 & 1.6 $\times$ 
10$^{-9}$ &  1.2 $\times$ 10$^{-6}$ & 1.62  & 112 \\
 G: NLTE/CNO+MEKAL            & 1.68     & 4.8  & 0.063  & & & 15.81 & 106 \\
 H: NLTE/NeOMg+MEKAL          & 1.69     & 5.0  & 0.068  & & & 10.73 & 105 \\ 
\hline
\end{tabular}
\end{table*}         
 
\section {Spectral analysis and interpretation}

In Paper I we made the working hypothesis that the supersoft flux
observed in November 1999
was entirely due to the central hot white dwarf remnant.
Neglecting the LECS flux below 0.8 keV
we simultaneously fitted  the LECS and MECS spectra 
with a {\sc mekal} model of thermal plasma
(included in the software package XSPEC, see Arnaud et al. 1986)
with parameters:  N(H)$\simeq$ 2 $\times$ 10$^{21}$
cm$^{-2}$, kT$\simeq$700 eV, and unabsorbed flux $\simeq$ 10$^{-12}$ erg
cm$^{-2}$ s$^{-1}$  (the reduced $\chi^2$ was $\approx$1.2).  
In this work we analyse instead the properties of the supersoft
X ray flux.
The spectral distribution observed for V382 Vel, shown in Fig.2 in the
range 0.1-1.1 keV, is strikingly different
from the one observed with the BeppoSAX LECS for three non-nova
supersoft X-ray sources:
 Cal 87 (Parmar et al. 1997), Cal 88 (Parmar et al. 1987)
and RX J0925.7-4758 (Hartmann et al. 1999).  The LECS 
spectra are shown in Fig. 7 of Hartmann et al.
(1999) and Fig. 3 of Parmar et al. (1997). V382 Vel
is much more luminous than the other observed sources, it appears to have
additional, harder spectral components and 
above all it is very luminous also 
in the very soft range, 0.1-0.2 keV. It is the only
object of this kind ever studied with the LECS. 

 We tried to fit the whole LECS and MECS spectra in the 0.1-10 keV
range adding the {\sc mekal} model used for the hard flux
 to {\it a)} a blackbody, {\it b)} a blackbody with absorption
edges (included in XSPEC), and {\it c)}  more detailed model atmospheres
 (see Hartmann \& Heise 1997, Hartmann et al. 1999). The latter models
were used to fit the non-nova supersoft X-ray sources and very
reasonable results were obtained.
We expected that the blackbody would not give a perfect fit
(since white dwarf atmospheres resemble blackbodies only
in first approximation) but we thought that it would
 provide an approximate estimate of the luminosity,
of the range of effective temperatures, and indicate whether 
absorption edges must be included.  Adding absorption edges to the black-body
(the most likely seemed C {\sc vi}  at 0.49 eV, but
we also tried N {\sc vi} at 0.55 keV and 0.67 keV, O {\sc viii} at 0.87 keV) 
did not improve the fit.
 
We also found that a reasonable fit to the supersoft
portion of the spectrum could not be obtained with
any of the atmospheric models we tested,  in Local
 Thermodynamical Equilibrium (hereafter, LTE), and in NON-LTE
(hereafter, NLTE), 
developed by Heise et al. (1994), Hartmann \& Heise (1997) and Hartmann et al. 
(1999).
For the NLTE case,  
we tested four  small grids of models with  log(g) between 8 and 9, 
with  four different set of abundances. The first set had 
cosmic abundances (developed primarily for galactic sources, hereafter
 NLTE-1 models),
in the second set of abundances the abundances were 
depleted to 0.25 the solar value (developed for LMC sources,
here after called NLTE-2 models), and in the third and fourth 
group of models the abundances were enhanced in C, N and O and 
in Ne, O and Mg, respectively.  These models include line opacities.
The two grids with enhanced abundances are still unpublished (Hartmann
2002, private communication).
 With either models in LTE and NLTE we could {\it not} obtain a good fit by,
for instance, decreasing log(g) and ``tuning'' the {\sl MEKAL}
 component accordingly.  In Table 1 we give the best fit parameters 
the NON-LTE   model atmospheres  with log(g)=8.5 and a {\sl MEKAL}
 bremsstrahlung component. The value log(g)=8.5 does not give
an acceptable value of $\chi^2$, but it is smaller than for other
values of log(g).
The only reason for testing also the NLTE-2 LMC-type models was that
we could not obtain a reasonable fit with the other available models
and  wanted to experiment with all the available grids. We concluded that
no model is acceptable: with the best NLTE-1 model, model A in the Table,
we obtain $\chi ^2$=4.5 for 105 degrees
of freedom. The fit to the data, shown in the upper part of Fig. 3, predicts  
counts at energies higher than 200 eV that differ from those observed
by more than 5\%, which is the likely uncertainty in the knowledge of the
LECS spectral response. Apart from what can be interpreted as an iron line
at $\approx$6.5 keV, at  low energies ($<$0.2 keV) the predicted counts
exceed those observed by at least 100\% (see Fig.2). However, the 
uncertainty in the low-energy LECS response in this range is estimated not 
to exceed $\simeq$20\% (Parmar et al. 1997).  

In a Chandra HRC-S+LETG
 observation of V382 Vel done in 2000 March by one of us,
 S. Starrfield,  the HRC-S+LETG spectrum, 
to be described in detail in a forthcoming
paper (Starrfield et al. 2002) is very different from another
nova which also appeared as a supersoft X-ray source,
V1494 Aql (N Aql 1999 no.2, see Starrfield et al. 2001).
The main difference is the lack of conspicuous continuum 
for V382 Vel in March 2000: it was instead an emission line spectrum, with many 
high ionization
emission lines in the supersoft range. These lines presumably 
have an origin in the ejected nebula.
The BeppoSAX instruments would have detected a blend of such
lines as a  featureless ``pseudocontinuum''. 
Emission  lines in the supersoft range may well have existed
even at the earlier epoch of our BeppoSAX observations. These lines
may have been due to 
 shock ionization in the ejecta,  rather than to  photoionization by 
 the central source.

 Even if the spectral resolution of the BeppoSAX LECS 
and MECS is not sufficient to detect lines, we tried to
determine which 
 emission lines may explain the observed X-ray spectrum and what
constraints we could derive on the level of the
continuum.  
In model C of Table 1 we added not only the bremsstrahlung continuum 
at the harder energy (kT=0.84 keV) and a line at $\approx$6.5 keV 
(which is necessary but is not in the range
where the bulk of the flux is emitted), but also one or more ``softer'' 
spectral features in emission.
As an experiment,  we obtained best fits with $\chi^2$=1.6--1.7 adding
such lines to both NLTE-1 and NLTE-2 models with log(g)=8.5 (see Table 1). 
  For the NLTE-1 model, in the best fit  the added  gaussian feature
has to be at 449 keV. It could probably  be C {\sc vi} 
(perhaps a blend of the C {\sc vi} triplet at 27-28 \AA). We tried
to add additional lines and improved the fit only marginally, although
another emission line at 243 eV may be present (probably the Fe {\sc xv} line
at 50.5 \AA), and
a narrow iron line at $\approx$6.4 keV is needed to explain the excess at 
this energy
(Model C).  This fit is shown in Fig. 3 (lower panel). 

With the NLTE-2 model we obtained model F with $\chi^2$=1.62 adding
a line (perhaps N {\sc vi}) at $\approx$490 keV, and again an iron
line for the excess at $\approx$6.5 keV. 
The  total  unabsorbed flux in the lines in these two models is
a negligible fraction of the total bolometric flux,  less than 1\%.
However,
the flux in the line at $\approx$450 or 490 keV would be 
about 30\% of the absorbed flux in the
BeppoSAX LECS range.  In model C,  
the bolometric luminosity at a distance of 2 kpc (Della Valle et al. 1999) 
is 6.8  $\times$ 10$^{38}$ erg s$^{-1}$. This value is
higher, but not much in excess of the model
 predictions for a $\approx$ 1 M$_\odot$ WD emitting at 
Eddington luminosity.

The main problem in determining the white dwarf parameters accurately
is the excess at low energy (kT$<$200 eV), which cannot be fit with one
or more narrow lines. Around 150 eV several transition
exist due to Fe, Si, Mg, and Ni, that could produce an intricate
pattern, must be heavily absorbed, and cannot 
be resolved with the spectral resolution of the BeppoSAX LECS.
By the time this nova was observed
with the Chandra LETG, it had become much less luminous and the 
spectral structure had definitely changed.   
We only remark that a complicated multi-temperature structure most likely
existed in the ejecta, and that continuum
and emission lines with different origins (white dwarf for the first and nebular
for the latter) can explain
the complicated spectrum we detected with the BeppoSAX LECS and MECS.
We rule out that our determination of effective temperature
and gravity of the white dwarf can be accurate 
if  nebular lines overlap with the white dwarf
continuum. It is only clear that at this stage the
atmosphere of the central source was still the dominating 
component of the X-ray flux.     

%
\begin{figure}
\centerline{\psfig{figure=lightcurve.ps,height=5.0 truecm,angle=-90}}
\centerline{\psfig{figure=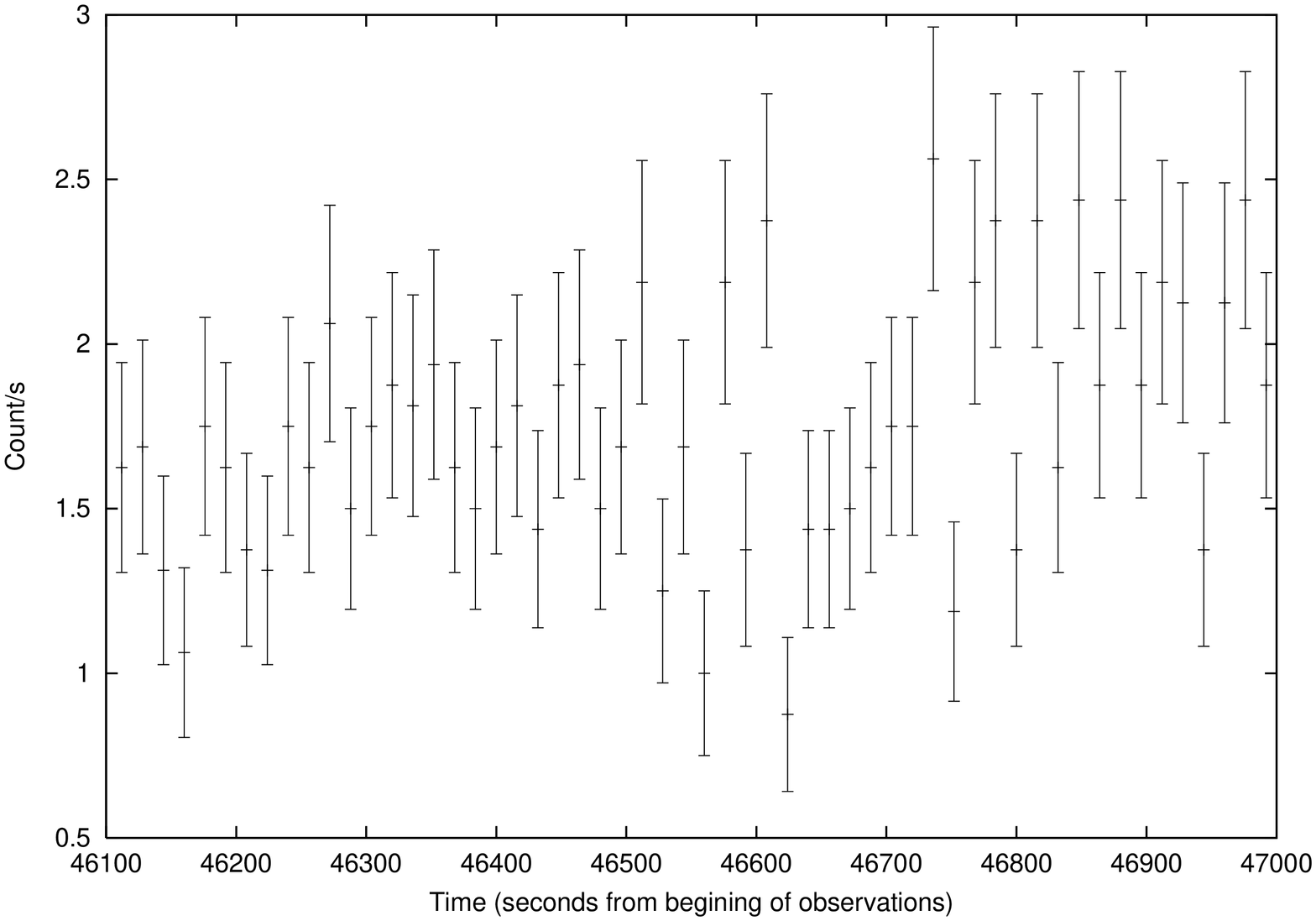,height=5. truecm,angle=-90}}
\centerline{\psfig{figure=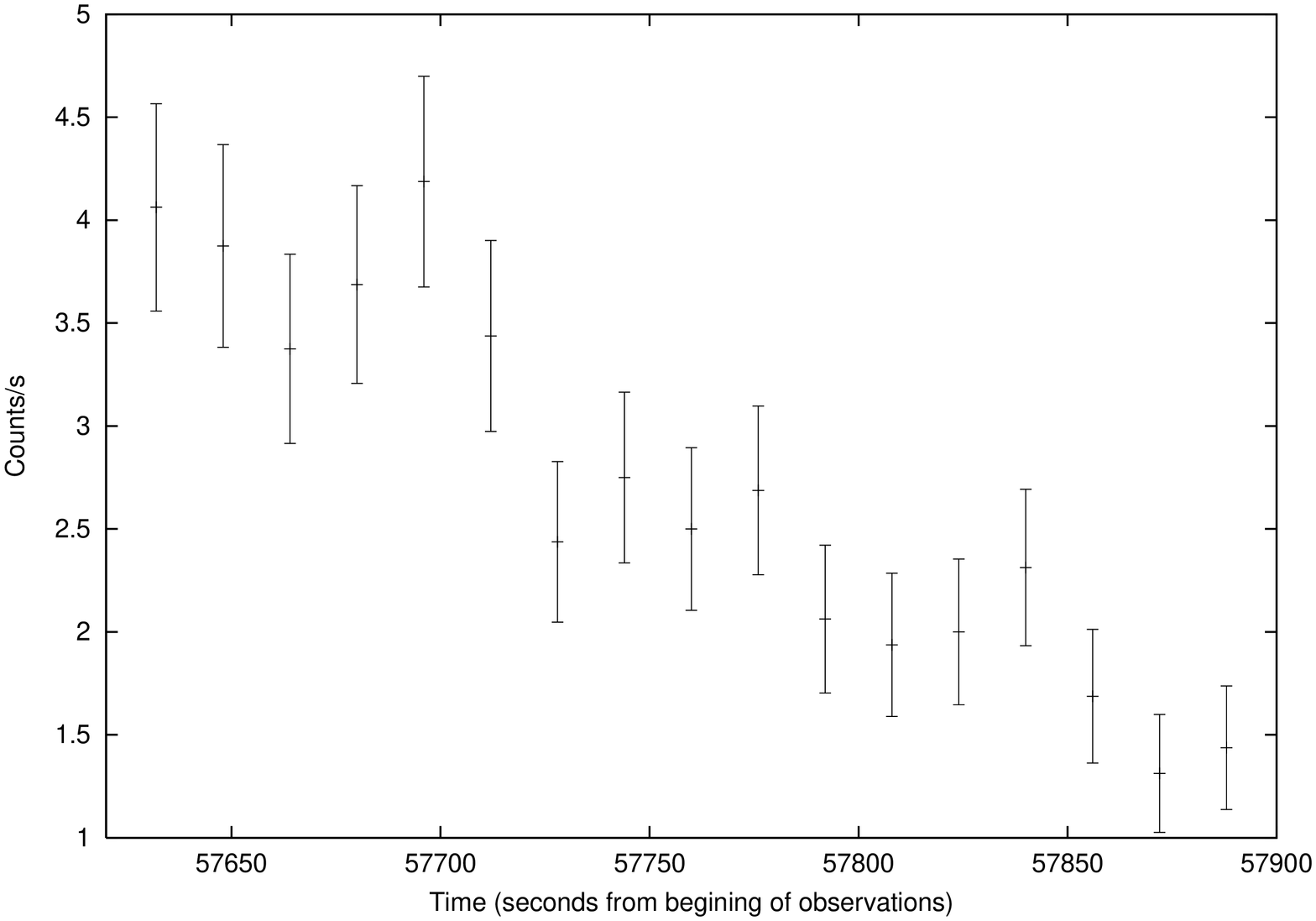,height=5. truecm,angle=-90}}
\caption{ The LECS lightcurve in the range 0.1-0.7 keV
with time bins of 100 s (top) and the 16 s binned lightcurve during the third before last
(second plot from the top) and the last
(bottom) of the 
observations done during the 16 hours from beginning to end of LECS exposure.
}  
\end{figure}
\begin{figure}
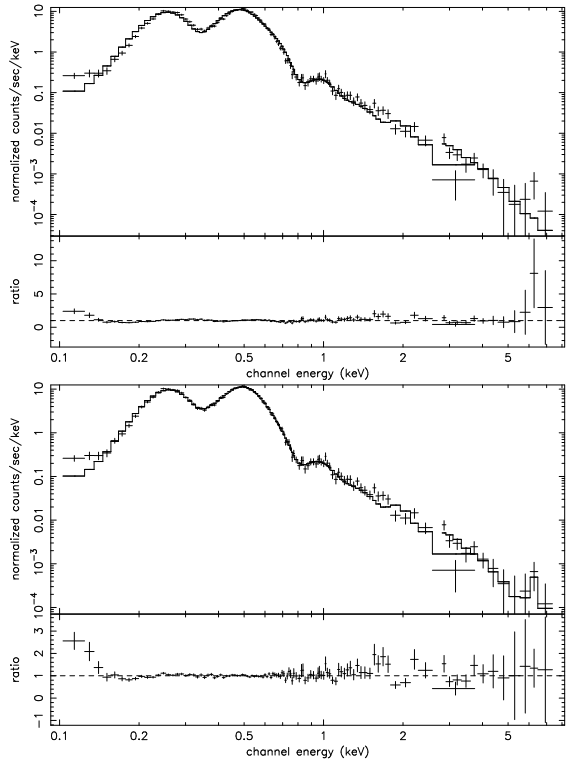

\centerline{\psfig{figure=Fig2a.ps,height=5.0truecm,angle=-90}}
\centerline{\psfig{figure=Fig2b.ps,height=5.0truecm,angle=-90}}
\caption{Spectra observed in the 0.1-8 keV range with
the BeppoSAX LECS and MECS in November 1999
 and (above) best fit with
model A ({\sc MEKAL} +  NLTE model atmosphere for log(g)=8.5 and
 cosmic abundances) and (below) with model C, which includes also three  
superimposed emission lines, at 0.243 keV, 0.449 keV and 6.4 keV). 
  The first fit is not acceptable,
and this is due to in a large portion to structured residuals
in the range 0.1-0.2 keV.  The second fit is overall much better,
but the residuals in the soft range are still large. The lower panel in
 each plot shows the residuals, as ratio
of observed over predicted counts per energy bin.}
\end{figure}       
\section{Conclusions}

The observation of V383 Vel 6 months after the outburst
revealed a very luminous supersoft X-ray source, comparable in luminosity
only with V1974 Cyg at maximum. The details of this observation raise new,
unexpected and very interesting, questions.
We observed irregular variability on a time scale of minutes in the supersoft
 flux of V382 Vel. 
The lack of energy dependence of the variability 
measured below 0.7 keV
seems to rule out the ejection of an obscuring clump, yet other
considerations seems to rule out orbital variability as well.

Moreover,
the BeppoSAX spectrum of V382 Vel in November 1999 appears much more complex
 than the expected
thermal continuum of a hot white dwarf plus a residual thermal
component from the nebula. We cannot justify this spectrum 
without invoking emission lines in the supersoft range
(which were  indeed observed shortly after this observation with Chandra),
so we suggest that the observed ``supersoft X-ray source'' in
V382 Vel is  characterized by unresolved narrow emission lines
superimposed on the atmospheric continuum. We 
found that even a contribution of
the lines of less than 1\% to the total bolometric flux
can significantly change the shape of the stellar continuum,
and make the task of determining white dwarf temperature and effective
gravity 
impossible with the resolution of the   BeppoSAX instruments.

However, we conclude that the bulk of the X-ray flux was still due to the 
atmospheric continuum and not to lines at this epoch,
unlike in the later observation performed with the Chandra
 LETG (Starrfield et al. 2002). 

Emission from classical novae in outburst can be quite
complex and different from one nova to another. 
The X-ray spectra of N LMC 1995
(Orio \& Greiner 1999) and of the recurrent nova U Sco (Kahabka et al.
1999) could be fit well with model
atmospheres, although
 U Sco required also a nebular component at higher energy. 
We caution however, that the spectral structure may
be as complex as the one observed for Cal 83 by Paerels et
al. (2001), when observed with higher resolution. 

The situation was more 
complex for V1974 Cyg, where a model atmosphere {\it and}
a hotter, thermal component were necessary to fit the spectrum.
The relative importance of the two components seemed to
 vary in each observation (see Balman et al. 1998)
and the interplay between them was rather
complicated, also due to the lower energy resolution
of the PSPC compared to the LECS (a factor 2.4 less) and limited energy
 range of the PSPC
(which could not cover the harder component well, at least at the
beginning).  We wonder whether the spectrum of V1974 Cyg 
also had superimposed nebular
emission lines that made it appear hotter than it was, because
 a lower effective temperature ($\leq$ 20 eV)
of the post-nova white dwarf atmosphere 
may explain a puzzling fact. An ionization nebula
was detected only in H$\alpha$ for N Cyg 1992 (Casalegno et al.
2000) while {\it other ionization lines} (indicating
a higher ionization potential) {\it were not present} in the nebula.
We speculate therefore that in  V1974 Cyg the white dwarf (Krautter et al.
1996, Balman et al. 1998) might have been {\it cooler} than
it appeared by fitting the ROSAT PSPC spectrum with just a two
component model. 

We note that even the spectrum of a non-nova super-soft
X-ray source, the Galactic MR Vel, shows non-atmospheric
emission lines, attributed to a wind from the source (Bearda et al. 2002). Instead,
in classical
novae in outburst emission lines in the soft X-ray range could be produced 
by shock ionization
 within the nebula. Shocked, X-ray emitting material 
seems to be present since the beginning of the 
outburst (Krautter et al. 1996, Orio et al. 2001b).
The BeppoSAX LECS and the ROSAT PSPC do not resolve narrow emission lines.     
 For novae, prominent nebular emission lines in the supersoft
X-ray energy range indicate interesting possibilities, specially
if they should be observed with Chandra or XMM-Newton in the future. 
We face new questions. Are these lines at times due to collisional
 excitation, do shocks occurs in the nova wind even many months
after the outburst? Should we consider
a line driven wind at different velocity colliding into the initial 
radiation driven wind? The nova theory must 
become more detailed and refined once the X-ray spectrum is known in
detail for a statistically meaningful sample of objects.
The gratings in the new X-ray observatories are opening new
and exciting possibilities for nova studies. 
 
\section*{Acknowledgments}
 
 We thank L. Piro, BeppoSAX Mission Scientist, for
support and W. Hartmann for providing the atmospheric models he developed. 
M. Orio is grateful also to another colleague, 
C. Markwardt, for computer support and useful advice. Finally,
we thank the referee, Koji Mukai.
This research has been supported by the Italian 
Space Agency (ASI) and by the UW College of Letters \& Sciences.
S. Starrfield gratefully acknowledges NSF and Chandra support to ASU.

\end{document}